\documentclass[conference]{IEEEtran}
\IEEEoverridecommandlockouts
\usepackage{cite}
\usepackage{amsmath,amssymb,amsfonts}
\usepackage{algorithmic}
\usepackage{graphicx}
\usepackage{textcomp}
\usepackage{xcolor}
\usepackage{multirow}
\usepackage{listings}
\usepackage{booktabs}
\usepackage{multirow}
\usepackage{graphicx}
\usepackage{makecell}
\usepackage{hyperref}
\usepackage{tcolorbox}
\usepackage[normalem]{ulem}
\usepackage{soul}
\usepackage{siunitx}

\def\BibTeX{{\rm B\kern-.05em{\sc i\kern-.025em b}\kern-.08em
    T\kern-.1667em\lower.7ex\hbox{E}\kern-.125emX}}

\def\BibTeX{{\rm B\kern-.05em{\sc i\kern-.025em b}\kern-.08em
    T\kern-.1667em\lower.7ex\hbox{E}\kern-.125emX}}

\newcommand{\RqOne}{What kinds of coinciding contributions occur during the mitigation period?}
\newcommand{\RqTwo}{How do developers distribute coinciding contributions during the mitigation period?}
\newcommand{\RqThree}{How related are coinciding contributions to the vulnerability?}

\makeatletter
    \newcommand{\linebreakand}{%
      \end{@IEEEauthorhalign}
      \hfill\mbox{}\par
      \mbox{}\hfill\begin{@IEEEauthorhalign}
    }
\makeatother

\definecolor{applegreen}{rgb}{0.55, 0.71, 0.0}
\definecolor{chestnut}{rgb}{0.8, 0.36, 0.36}
\definecolor{gray(x11gray)}{rgb}{0.75, 0.75, 0.75}
\definecolor{mygray}{RGB}{128,128,128}

\begin{document}

\title{Characterising Contributions that Coincide \\ with Vulnerability Mitigation in NPM Libraries 
}


\author{\IEEEauthorblockN{Ruksit Rojpaisarnkit, Hathaichanok Damrongsiri, \\ Christoph Treude*, Ali Ouni$^{\dagger}$, Raula Gaikovina Kula}
\textit{Nara Institute of Science and Technology, Japan}\\
\textit{Singapore Management University, Singapore*}\\
\textit{ETS Montreal, University of Quebec, Canada$^{\dagger}$}\\

\{rojpaisarnkit.ruksit.rn1, damrongsiri.hathaichanok.db5, raula-k\}@is.naist.jp \\ ctreude@smu.edu.sg, ouniaali@gmail.com}









\maketitle

\begin{abstract}
With the urgent need to secure supply chains among Open Source libraries, attention has focused on mitigating vulnerabilities detected in these libraries. Although awareness has improved recently, most studies still report delays in the mitigation process. This suggests that developers still have to deal with other contributions that occur during the period of fixing vulnerabilities, such as coinciding Pull Requests (PRs) and Issues, yet the impact of these contributions remains unclear. To characterize these contributions, we conducted a mixed-method empirical study to analyze NPM GitHub projects affected by 554 different vulnerability advisories, mining a total of 4,699 coinciding PRs and Issues. 
We believe that tool development and improved workload management for developers have the potential to create a more efficient and effective vulnerability mitigation process.
\end{abstract}



\section{Introduction}
Developers are increasingly becoming aware of the ever-growing threat of security vulnerabilities in their third-party dependencies \cite{Web:octverse}. 
Although such awareness has improved over recent times, most studies still report lags in the update in response to third-party components.
For instance, a plethora of work \cite{Kula:2017,Bodin:EMSE2021,Alfadel:SANER2021,Decan:2018,Cox-ICSE2015,ZeroualiArxiv2021} report lags in the update of security updates and has generally been the case for any dependency update in the ecosystem.
Developers often cite roles, being undermanned, and the migration effort needed to mitigate the risk \cite{Bogart:2015,BogartFSE2016}.
Efforts have also been driven by the Industry. 
For instance, the Alpha and Omega Project specifically aims to fast-track the most critical open source projects to improve their security postures~\cite{AlphaOme74:online}.

Due to their open-source nature, the key risk of using these libraries is that contributors (including maintainers) may become overwhelmed with the maintenance of their volunteer work.
For instance, recently the maintainers of
a vulnerable library expressed their frustration when maintaining a library in
a tweet, \textit{`... maintainers have been working sleeplessly on mitigation measures; fixes, docs, CVE, replies to inquiries, etc. Yet nothing is stopping people from bashing us, for work we aren't paid for, for a feature we all dislike yet needed to keep due to backward compatibility concerns...' \footnote{https://twitter.com/yazicivo/status/1469349956880408583}.}

Our aim is to study how these coinciding contributions impact npm packages.
We use the term coinciding to describe contributions that were submitted and merged by projects while being aware of an existing vulnerability in the project.
To achieve our goal, we have derived the following research questions:


    \noindent \textbf{RQ1:} \textbf{\RqOne}\\
    \underline{\textit{Motivation:}} The motivation of RQ1 is to understand the kinds of activities developers are faced with during the mitigation period.
    By answering this research question, we should be able to understand what kinds of maintenance activities exist.\\
    \underline{\textit{Results:}}
    Results indicate that the vulnerability mitigation period involved attending to bugs and features. Furthermore, we find that except for other issues, most coinciding contributions are likely to be merged into the codebase. \\
    \noindent \textbf{RQ2:} \textbf{\RqTwo}\\
    \underline{\textit{Motivation:}} The motivation of RQ2 is to understand how much of the vulnerability mitigation period is spent resolving coinciding contributions. 
    This insight will help us understand the extent to which coinciding contributions are working in parallel with the vulnerability fix.\\
    \underline{\textit{Results:}}
    Developers spend on average around 50\% to 60\% of the vulnerability mitigation period attending to coinciding contributions. This includes both merged and abandoned PRs and Issues. \\
    \noindent \textbf{RQ3:} \textbf{\RqThree}\\
    \underline{\textit{Motivation:}}
    Anecdotal evidence suggests that these volunteer developers are being overwhelmed with their workload. Hence, our motivation for RQ3 is to gain a deeper understanding of the extent to which these coinciding contributions are actually related to the vulnerability.
    \\
    \underline{\textit{Results:}}
    On one hand, results indicate that a majority of coinciding contributions are not related to the vulnerability.
On the other hand, the developers who are involved in mitigating the vulnerability are also involved in 30\% of the coinciding contributions.

\begin{table}[]
\centering
\caption{Statistics of the Data Collection}
\label{tab:dataset_statistic}
\scalebox{1}{
\begin{tabular}{@{}lr@{}}
\toprule 
\multicolumn{2}{c}{\textbf{Collecting Advisories} } \\ \midrule
snapshot period &  1st-Oct-2017 to 1st-April-2023 \\
\# Collected GitHub Advisories & 11,879 Advisories \\ 
\# Advisories after NPM filter & 2,933  Advisories\\ \midrule
\# Advisories after GitHub filter & 554  Advisories \\ \midrule
\multicolumn{2}{c}{\textbf{Matching Affected NPM Libraries} } \\ \midrule
\# Matching GitHub Repos & 348 \\ 
\# Total Contributions & 1,225,489 \\
-- \# PRs & 402,175 \\
-- \# Issues & 823,314 \\\midrule
\multicolumn{2}{c}{\textbf{Contributions that coincide} } \\ \midrule
\# Total Contributions & 4,706 \\
-- \# Coinciding PRs &  2,159 \\
-- \# Coinciding Issues & 2,547 \\
\bottomrule
\end{tabular}
}
\end{table}


\section{Data Preparation}
In this section, we present our data collection sources and processing. 
Figure \ref{fig:architecture} describes an overview of our data collection methodology. 
We now discuss the details below.

\subsection{Collecting Vulnerabilities from the NPM Ecosystem}
We decided to use the NPM ecosystem as our case study.
NPM is one of the largest package ecosystem, and has been the focus of recent studies \cite{Decan:emse2019}. 

\paragraph{Data Sources}
We collect our data from two sources.
The first is the GitHub Advisory Database. 
Since we chose the NPM ecosystem, the GitHub Advisory Database collected vulnerabilities from both the National Vulnerability Databases and the npm Securities Database. 
As we evaluted contributions to GitHub projects, it was more likely that the Advisories would also contain explicit links to any PRs and Issues that mitigates the vulnerability.


\paragraph{Extracting Advisories}
As shown in Table \ref{tab:dataset_statistic}, we first downloaded all advisories from October 1st 2017 to April 1st. 
The reason from this time period is the the GitHub Advisories had its first inception in 2017. 
From these 11,879 Advisories, we then filtered out 554 Reports that affected NPM libraries. 

\paragraph{Extract and Identifying Coinciding Contributions}
From the 348 advisories, we were able to match them with 348 NPM projects hosted in GitHub.
In this study, we refer to contributions as any PR and Issue that were created for those vulnerable projects.
We then extracted all PRs and Issues from these projects, ending up with a total of 402,175 PRs and 823,314 issues from these 348 repositories.

Figure \ref{fig:coinciding} depicts the two types of coinciding contributions.
We define a contribution that coincides with a vulnerability mitigation, which is any contribution that is closed during
We define the following concepts:
\begin{itemize}
    \item \textit{Mitigation Period (\# days).} This is the period by which the vulnerability is first created, to when it is closed. In our figure, we can see that the vulnerability is created at time \textit{t2}, and resolved at \textit{t6}. The unit of measurement is in days.
    \item \textit{Coinciding contributions.} We identify two types of contributions that coincided with the mitigation period. Generally, we identify any contribution that is closed before the vulnerability is mitigated. As shown in the Figure, this is any PR or Issue that was closed before time t6.
\end{itemize}
Hence, from the 348 GitHub repositories we were able to filter out 2,159 PRs and 2,547 Issues that met this criteria.
Note that all research questions (RQ1, RQ2, and RQ3) use this quantitative dataset.
Since RQ3 required a deeper understanding of the PRs and Issues, we also conducted a statistical sampling of our dataset. 
Following prior work, we then took a sample with a configuration that includes 95\% of confidence level and confidence intervals equal to 5 from 2,159 PRs and 2,547 Issues.
This left us with 326 PRs and 334 Issues for manual analysis.
The dataset of the experiments is available on Zenodo, at \url{https://doi.org/10.5281/zenodo.10978818}.



\section{Findings}
In this section, we present our findings, which includes the approach, reported results, and then finally answering each research question.

\subsection{\textbf{(RQ1) \RqOne}}

\paragraph{\underline{\textbf{Approach}}}
To answer RQ1, we conduct a quantitative analysis of the collected coinciding contributions. 
We classify each contribution using the codes based on the taxonomy of \cite{Subramanian:2022}:
\begin{itemize}
    \item \textit{Documentation}: changes and additions made to documentation files such as READMEs and/or comments explaining code. Note that only commits where the majority change is documentation are classified as documentation changes.
    \item \textit{Feature}: adding new functionality/features to the project.
    \item \textit{Bug}: fixing unexpected behavior in code.
    \item \textit{Refactoring}: restructuring code to make it more understandable/readable and/or conform to coding standards.  
    \item \textit{Test case}: adding test cases and/or adding code to facilitate testing.
    \item \textit{Other}: anything that does not fall in these categories.
\end{itemize}

In our approach, we used a semi-automatic method, by using a combination of keyword searching, based on a sample of contributions (30 PRs and Issues) that three authors manually discussed over a round table. 
For each contribution, we first classified based on the keyword.
The resulting keywords are shown in Table \ref{tab:automatic_keyword}.

\begin{table}[]
\centering
\caption{Keywords for each category}
\label{tab:automatic_keyword}
\scalebox{1}{
\begin{tabular}{c|p{4cm}}
\toprule 
\multicolumn{1}{c}{\textbf{Type}} & \multicolumn{1}{c}{\textbf{Classification keywords}}\\ \midrule
Bug & avoid, fix, resolve, close, bug, solve, solution, issue, fixing \\
Feature & integrate, add, feat, update, upgrade, support, dependencies, feature, improve, version, automate, compatibility, bundle, improvement, bump\\
Test & test case, test, unit test, CI, continuous integration \\
Refactoring & remove, unnecessary, refactor, performance, optimize\\
Documentation & documentation, doc\\
\bottomrule
\end{tabular}
}
\end{table}

\begin{figure}[!]
    \centering
    \includegraphics[width=1\linewidth]{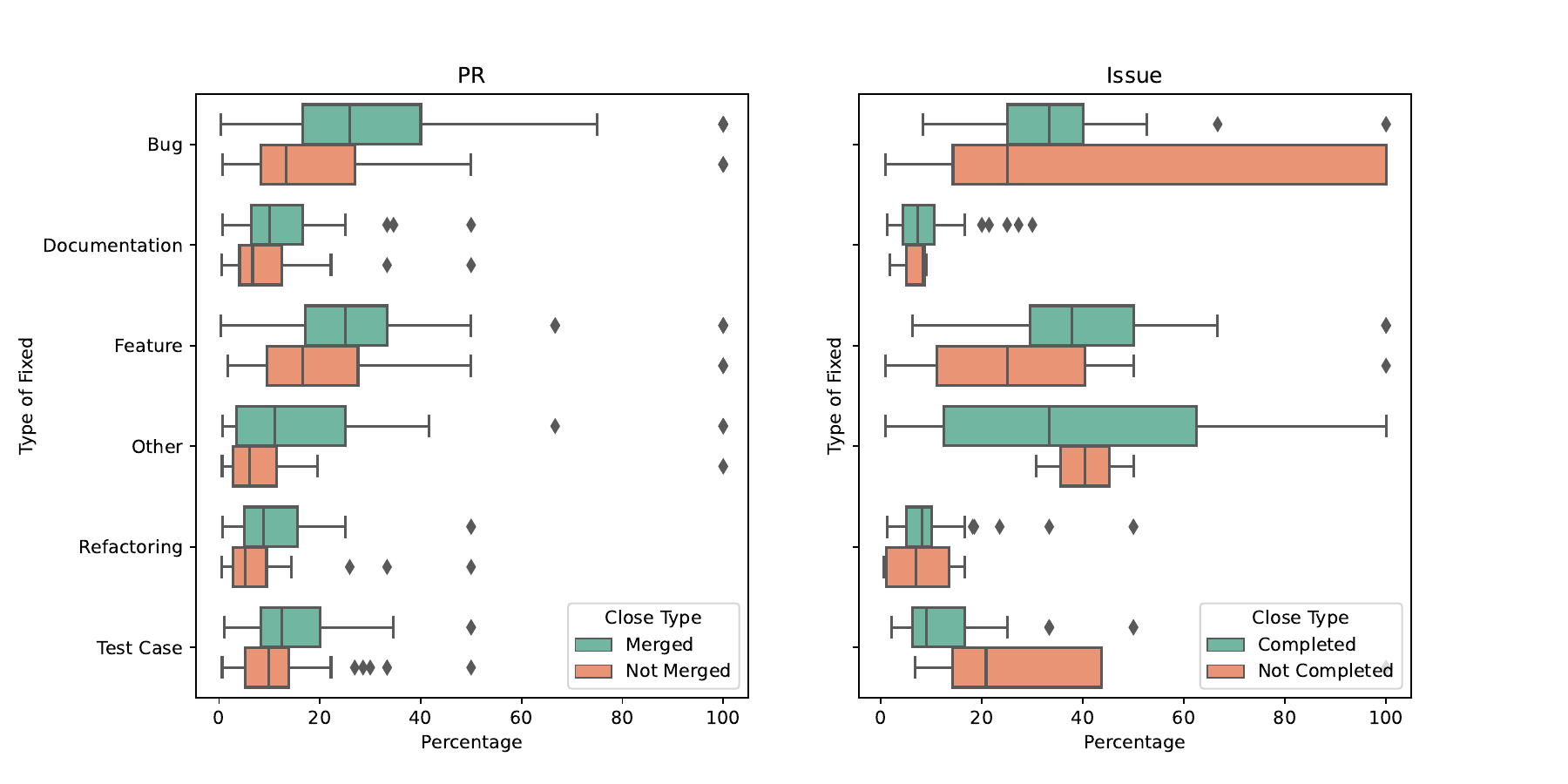}
    \caption{Kinds of coinciding contributions}
    \label{fig:rq1_kinds}
\end{figure}

\begin{table*}[]
\centering
\caption{p-value $<$ 0.01, PR cliff delta value}
\label{tab:rq1_correlation}
\scalebox{0.9}{
\begin{tabular}{@{}l|l|llllll@{}}
\toprule
Type & Category & Bug & Documentation & Feature & Refactoring & Test Case & Other \\ \midrule
\multirow{6}{*}{PR} & Bug & (N/A) & \textcolor{purple}{(0.65, large)} & \textcolor{teal}{(0.02, negligible)} & \textcolor{purple}{(0.69, large)} & \textcolor{purple}{(0.55, large)} & \textcolor{purple}{(0.63, large)} \\
 & Documentation & \textcolor{purple}{(-0.65, large)} & (N/A) & \textcolor{purple}{(-0.62, large)} & \textcolor{teal}{(0.05, negligible)} & \textcolor{violet}{(-0.19, small)} & \textcolor{teal}{(0.14, negligible)} \\
 & Feature & \textcolor{teal}{(-0.02, negligible)} & \textcolor{purple}{(0.62, large)} & (N/A) & \textcolor{purple}{(0.65, large)} & \textcolor{purple}{(0.53, large)} & \textcolor{purple}{(0.59, large)} \\
 & Refactoring & \textcolor{purple}{(0.69, large)} & \textcolor{teal}{(0.05, negligible)} & \textcolor{purple}{(0.65, large)} & (N/A) & \textcolor{violet}{(-0.26, small)} & \textcolor{teal}{(0.11, negligible)} \\
 & Test Case & \textcolor{purple}{(-0.55, large)} & \textcolor{violet}{(0.19, small)} & \textcolor{purple}{(-0.53, large)} & \textcolor{violet}{(0.26, small)} & (N/A) & \textcolor{violet}{(0.29, small)} \\
 & Other & \textcolor{purple}{(-0.63, large)} & \textcolor{teal}{(-0.14, negligible)} & \textcolor{purple}{(-0.59, large)} & \textcolor{teal}{(-0.11, negligible)} & \textcolor{violet}{(-0.29, small)} & (N/A) \\ \midrule
\multirow{6}{*}{Issue} & Bug & (N/A) & \textcolor{purple}{(0.61, large)} & \textcolor{violet}{(-0.17, small)} & \textcolor{purple}{(0.58, large)} & \textcolor{purple}{(0.53, large)} & \textcolor{teal}{(0.10, negligible)} \\
 & Documentation & \textcolor{purple}{(-0.61, large)} & (N/A) & \textcolor{purple}{(-0.68, large)} & \textcolor{teal}{(-0.06, negligible)} & \textcolor{teal}{(-0.14, negligible)} & \textcolor{brown}{(-0.45, medium)} \\
 & Feature & \textcolor{violet}{(0.17, small)} & \textcolor{purple}{(0.68, large)} & (N/A) & \textcolor{purple}{(0.65, large)} & \textcolor{purple}{(0.61, large)} & \textcolor{violet}{(0.18, small)} \\
 & Refactoring & \textcolor{purple}{(0.58, large)} & \textcolor{teal}{(-0.06, negligible)} & \textcolor{purple}{(0.65, large)} & (N/A) & \textcolor{teal}{(-0.08, negligible)} & \textcolor{brown}{(0.41, medium)} \\
 & Test Case & \textcolor{purple}{(0.53, large)} & \textcolor{teal}{(0.14, negligible)} & \textcolor{purple}{(-0.61, large)} & \textcolor{teal}{(0.08, negligible)} & (N/A) & \textcolor{brown}{(-0.34, medium)} \\
 & Other & \textcolor{teal}{(-0.10, negligible)} & \textcolor{brown}{(0.45, medium)} & \textcolor{violet}{(-0.18, small)} & \textcolor{brown}{(0.41, medium)} & \textcolor{brown}{(0.34, medium)} & (N/A) \\
\bottomrule
\end{tabular}
}
\end{table*}

For analysis and statistical significance testing, we would like to test the null hypothesis that ‘the percentage of each coinciding contribution type between advisories is the same’ then whether or not coinciding code changes are prevalent, and whether coinciding code changes are more likely to be accepted by the maintainers.
Hence, we apply the Kruskal-Wallis H-test (Kruskal and Wallis, 1952), which
is a non-parametric statistical test to use when comparing more than two groups. If there is significance, then we also measure the effect size using Cliff’s $\delta$, a non-parametric effect size measure. Effect size is analyzed as follows: 
\textit{(1) |$\delta$| $<$ 0.147 as Negligible}
\textit{(2) 0.147 $\leq$ |$\delta$| $<$ 0.33 as Small}
\textit{(3) 0.33 $\leq$ |$\delta$| $<$ 0.474 as Medium}
\textit{(4) 0.474 $\leq$ |$\delta$| as Large} For statistical power, 
We use the cliffsDelta \footnote{\url{https://github.com/neilernst/cliffsDelta}} 
package to analyze Cliff’s $\delta$.

\paragraph{\uline{\textbf{Results}}}

Figure~\ref{fig:rq1_kinds} shows the results of our analysis.
We find that a significant number of PRs and Issues can be characterized as either 30.97\% Bugs or 33.5\% Features. 
Another interesting observation is that statistically we show that most of these coinciding changes are indeed merged into the codebase.
Visually, for Issues, we do not see this case, with many of the issues labeled as not completed when finally closed. 

Table\ref{tab:rq1_correlation} shows the statistical tests to test our null hypothesis on 'the percentage of each coinciding contribution type between advisories is the same' is rejected (i.e. p-value $<$ 0.001). From Cliff’s $\delta$ effect size, Feature and Bug are negligible and significantly larger than other types for PR. In addition, Documentation, Refactoring, Test case, and Other are small and negligible. On another hand, Cliff’s $\delta$ effect size shows that instead of Bug and Feature, Other are also larger than others while the effect size is small and negligible compare with Feature and Bug respectively for issues.

\subsection{\textbf{(RQ2) \RqTwo}}

\paragraph{\uline{\textbf{Approach}}}
To answer RQ2, we take a quantitative approach to calculate whether or not developers are quickly resolving coinciding contributions.
Our hypothesis is that developers may seek to complete coinciding contributions earlier in the mitigation period. A competing and alternative hypothesis is that developers might work in parallel and resolved all contributions in the same manner.

\begin{figure}[!]
    \centering
    \includegraphics[width=0.95\linewidth]{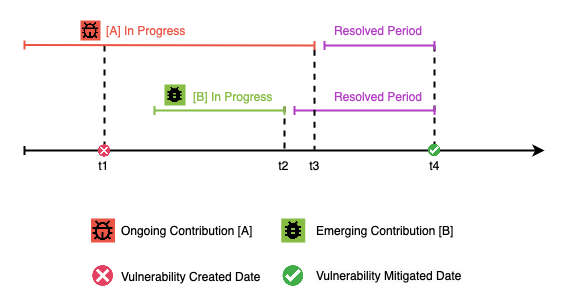}
    \caption{Terminology of Coinciding Contributions}
    \label{fig:terminology}
\end{figure}

Figure \ref{fig:terminology} describes our method to calculate the resolving time for each coinciding contribution.
Shown in purple, the resolved period is calculated as the difference in days between when the vulnerability was mitigated, from each of the types of coinciding contributions. 
Hence, we define the following terminology:
\begin{itemize}
    \item \textit{Resolve-time(\#days):} The resolve-time is defined as the time between when the coinciding contribution was closed and the when the vulnerability was closed. In this example, show the resolving time for the ongoing contribution [Case A] as being from t3 to t4, and the emerging contribution [Case B] as being from t2 to t4.
    \item \textit{Coincide-freed (\%):} Using the resolve-time, we can now calculate the proportion of the mitigation that is free of the coinciding contribution. Concretely, this is calculated as the resolve-time divided by the mitigation period. In our example, the Coincide-free-percent for Case B is calculated as $\frac{diff(t2,t4)}{diff(t1,t4)}$ where $diff$ is the time difference between two dates. Note that a higher percentage indicates that developers are more likely mitigate vulnerability concurrently to coinciding contributions.
\end{itemize}

For our analysis, we would like to test two hypothesis. 
The first is to test the assumption that issues are more likely freed up when compared to PRs.

Similar to RQ1, to assess the statistical difference between related PRs and other PRs, we apply the Mann-Whitney U test \cite{mann1947test}, which is a non-parametric test that is used to compare two sample means come from the same population. We also measure the effect size using Cliff’s $\delta$, a non-parametric effect size measure.
Effect size is analyzed as follows: 
(1) |$\delta$| $<$ 0.147 as Negligible, 
(2) 0.147 $\leq$ |$\delta$| $<$ 0.33 as Small, 
(3) 0.33 $\leq$ |$\delta$| $<$ 0.474 as Medium, or 
(4) 0.474 $\leq$ |$\delta$| as Large. We use the cliffsDelta 
package to analyze Cliff’s $\delta$.
Overall, we found a statistical difference between PRs created before and after the disclosure, confirming that developers had increased their development activities in general (\textit{p}-value < 0.01, Cliff’s $\delta$ is Large). 

\paragraph{\uline{\textbf{Results}}}


\begin{figure}[t]
    \centering
    \includegraphics[width=0.8\linewidth]{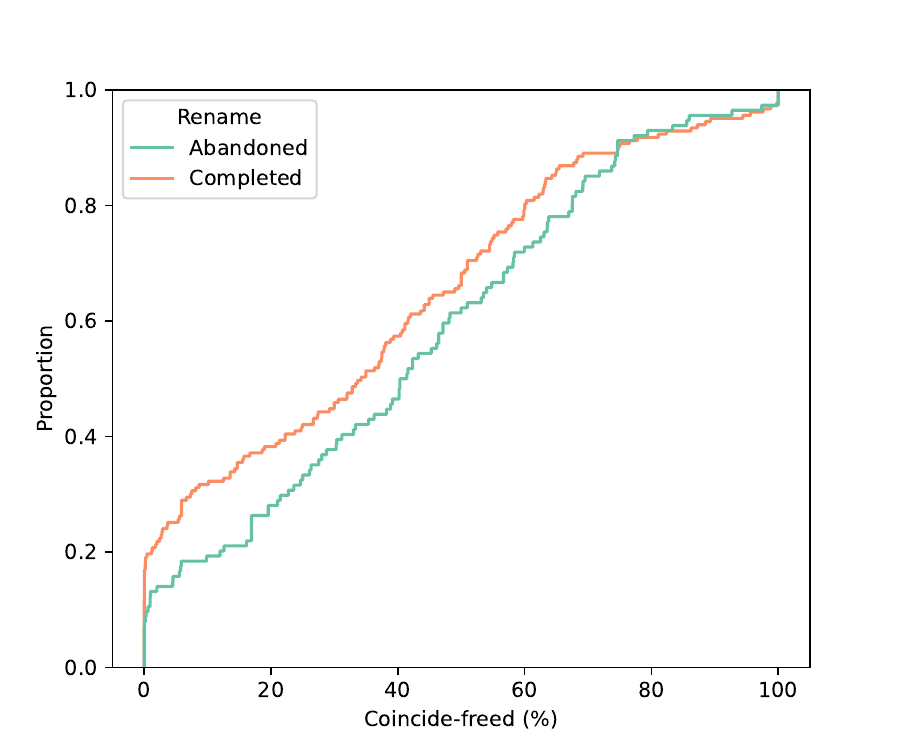}
    \caption{Distribution of the proportion of time used for closing coinciding contribution.}
    \label{fig:rq2_1}
\end{figure}

From our experiment, we found that maintainers use 45.89\% of the vulnerability mitigation period on average to work on the coinciding contribution while the vulnerability still exists. In addition, the maximum and the minimum amount of time is 100\% and 0\% respectively, which means some coinciding contributions have been done on the same day on the vulnerability created date and the vulnerability mitigated date.

\subsection{\textbf{(RQ3) \RqThree}}

\paragraph{\textbf{\uline{Approach}}}
To answer RQ3, we conducted a mixed-method to understand how related the coinciding contributions are to the vulnerability.
Hence, we conceptualize relatedness from two perspectives. We outline these analysis below:

\textbf{Analysis 1: \textit{Relatedness by Maintainer}}. The first definition of relatedness is based on whether the same maintainer that is assigned to vulnerability, also is assigned to the coinciding contribution.
To do so, for each vulnerability we first identify which maintainers are assigned to the contribution.
We then systematically classify whether or not a coinciding contribution includes these assigned maintainers.
Our assumption is that a higher proportion of coinciding contributions may be related by the Maintainer.
To achieve this, we used a quantitative experiment to classify what percentage of maintainers were involved in both activities. We analyzed the merged\_by field of pull requests and closed\_by field of issues for both the coinciding contributions and the vulnerability, comparing the login names between the two to determine whether or not they were the same. 

To identify the effect of this types of relatedness, we can also measure whether or not the relatedness is correlated with the coincide-freed (defined in RQ2).
Our hypothesis is that maintainers are more likely to freed up related coincided contributions compared to those that are not.

\textbf{Analysis 2: \textit{Relatedness by Security Mentions}}. The second definition of relatedness is based on whether the coinciding contribution actually mentions the security vulnerability in its description, title, or the comments that are made on the contribution. 
To do so, we conduct a manual analysis on a sample of 326 pull requests and 334 issues.
Specifically, three authors conducted a round-table coding, examining  the title, description, comments, and related commits for each pull request and issue.
The authors carefully discussed any mention of the vulnerability or updates to dependency versions. 

To characterise these relatednesses, we also classify the relatedness with the kinds of code changes (defined in RQ1).
Our hypothesis is that specific types of coinciding contributions might be related to the security vulnerability.

\paragraph{\textbf{\uline{Results}}}
Result indicates that 37.99\% of coinciding pull requests and 
20.18\% of coinciding issues have been done by the same maintainer who fixed the vulnerability. Moreover, there are only 2.2\% of the samples that have been classified related to the vulnerability. 67.8\% of the sample is not related to the vulnerability and also not related to any updating dependency for their library. Most of the coinciding contributions for PR only fixed some bugs, adding the feature and other contributions that are not related to the vulnerability. On another hand, most of the issues related to answering the question, reporting some bugs, and giving instructions.

\section{Implications and Challenges}

For Practitioner, our results indicate that during the vulnerability period, practitioners did not prioritize fixing the vulnerability as the first priority. Briefly, there are numerous contributions have been done during the vulnerability period which are not related to the vulnerability. According to our experiments, those contributions are, for example fixing bugs, and introducing new features and others, which not related to the vulnerability. Moreover, it has been contributed by the same person who takes responsibility for fixing the vulnerability. Therefore, practitioners should be aware of how to prioritize the activity during the vulnerability period which in practice, the vulnerability should be fixed as fast as possible.

For Researcher, according to results that illustrate the activity of the practitioner during the vulnerability period, it indicate that there is a gap to study on how practitioners should manipulate their tasks and workload during different situations. Especially in specific situations such as a vulnerability period. Since the previous research mostly focuses on how to detect vulnerability, the researcher can look more into the best practices to deal with the vulnerability in the perspective of manipulating the workload.

\section{Threats to Validity}
External Validity is concerned with our ability to generalize based on our results.
We only conduct an empirical study on the NPM ecosystem.
As such, observations based on this study may not generalize to other software ecosystems.
However, the NPM ecosystem is nowadays regarded as the popular software ecosystem involved with a large number of libraries and contributors.

Construct Validity refers to the relation between theory and observation.
Firstly, we collected the data from only one source. Thus, there might be some vulnerabilities that our dataset does not contain in our source.
We believe that it may not have a large impact on our findings in this paper since our experiment tends to focus on pull requests and issues
Secondly, in our qualitative analysis of contribution taxonomy (i.e., feature, bug, refactoring, etc.), the keywords used to classify types of contribution are brought from previous related work that might not coverage. 
To mitigate this threat, we perform the round table discussion between all authors with the sample of classification to verify the correctness and coverage of the keyword classification.

Internal Validity is the approximate truth about inferences about cause-effect or causal relationships. 
Two related threats are summarized.
The first threat is related to the metrics that we selected and measured to identify the relationship between groups of taxonomy.
For instance, when we investigate the amount of contribution type during the vulnerability period, the number of pull requests and issues may not directly indicate the workload of developers.
Meanwhile, apart from the number of pull requests and issues, several confounding factors may affect the workload of the developer, but our goal in this stage is to shed light on the role of coinciding contribution.
The second threat concern is the selection of statistical tests.
To validate our hypotheses, we adopted different hypothesis tests (Kruskal-Wallis H-test and Cliff’s $\delta$) to identify the characteristics of the target data (whether independent or not).

\section{Related Work}
\label{sec:related}

\textbf{Security Vulnerability.} The prevalence of third-party libraries in modern software development creates a ripple effect, where vulnerabilities in one library can impact the security of numerous dependent applications. This interconnectedness forms a software ecosystem susceptible to chain security problems. Recent studies have explored dependency networks in different aspects. Some studies investigated the structure and evolution of the dependency networks and revealed their issues such as dependency hell and technical lag~\cite{Decan:2017, Kikas:2017, Zerouali:ICSR2018}. Other several Studies reveal that software often lags behind in updating libraries, leaving them vulnerable. \cite{Bavota2015, Hora2015, robbes2012, sawant2016} Conversely, \cite{mirhosseini} suggests that pull requests with badge notifications can expedite the update process. To further explore this issue, this research investigates contributor activities during the vulnerability fixing period to identify additional factors contributing to update delays.

\textbf{Mining Repositories.} Numerous studies leverage data mining techniques to extract valuable insights from software repositories and ecosystems. Within the context of the NPM ecosystem, researchers have explored various data sources embedded within different files. For instance, \cite{Alhazmi:2007, Chowdhury:2011} focused on analyzing source code within commits, while \cite{Mirhosseini:2017} and \cite{Wittern_MSR2016} extracted data from the package.json file. Additionally, vulnerability analysis often integrates data from standardized resources such as the Common Weakness Enumeration (CWE). This work applies a similar technique to collect data from the GitHub repository.

\section{Conclusion and Future Work}

We highlight two takeaways and the research outlook as follows.
\paragraph{Rethinking How Developers Fix Severe Threats}
Our key takeaway message is that when it comes to a critical vulnerability, instead of dropping everything, developers tend to increase their activities, resolving all pending PRs and issues before attending to the fix.
We make the argument that it is equally important to assist developers with resolving pending issues, as this affects how quick, and the resources needed to seemingly fix a single vulnerability.
For instance, it was reported on social media that the Log4J support team would have welcomed assistance \textit{"...Log4j maintainers have been working sleeplessly on mitigation measures; fixes, docs, CVE, replies to inquiries, etc...Yet nothing is stopping people to bash us, for work we aren't paid for, for a feature we all dislike yet needed to keep due to backward compatibility concerns."} 
\cite{web:Twitter2021}
We argue that research into understanding the prerequisites for fixing a vulnerability is needed.
A concrete example is tool support to inform developers of pending issues and PRs that may or may not affect a library.
This change in mindset promotes best practices to maintaining third-party interfaces to the code (Application Programming Interfaces).
How this mindset is perceived by developers and its barriers is deemed as future work.

\paragraph{Recommending Information Needs to Fix a Severe Threat}
Our results motivate the need for developers to readily access information on the vulnerability. 
Although these are existing global information hubs like the vulnerability database, and generic blogs.
Our immediate research direction is to use crowd-sourcing to search and recommend information across different projects that are resolving the same vulnerability. 

\section*{Acknowledgement}
This work is supported by the Japanese Society for the Promotion of Science (JSPS) KAKENHI Grant Number JP20H05706.

\vspace{12pt}

\bibliographystyle{IEEEtran}
\bibliography{references}
\end{document}